\documentclass[%
 aip,
 amsmath,amssymb,
 reprint,%
]{revtex4-1}

\usepackage{graphicx}
\usepackage{dcolumn}
\usepackage{bm}

\usepackage[utf8]{inputenc}
\usepackage[T1]{fontenc}
\usepackage{mathptmx}
\usepackage{etoolbox}
\usepackage{xcolor}

\def\@email#1#2{%
 \endgroup
 \patchcmd{\titleblock@produce}
  {\frontmatter@RRAPformat}
  {\frontmatter@RRAPformat{\produce@RRAP{*#1\href{mailto:#2}{#2}}}\frontmatter@RRAPformat}
  {}{}
}%
\makeatother
\begin{document}

\preprint{AIP/123-QED}

\title[]{Non-Markovian exceptional points in waveguide quantum electrodynamics}
\author{Stefano Longhi}
 \email{stefano.longhi@polimi.it}
 \altaffiliation[Also at ]{IFISC (UIB-CSIC), Instituto de Fisica Interdisciplinar y Sistemas Complejos - Palma de Mallorca, Spain}
\affiliation{ Dipartimento di Fisica, Politecnico di Milano, Piazza L. da Vinci 32, I-20133 Milano, Italy}

\date{\today}
\begin{abstract}
Spontaneous emission of a quantum emitter, such as an excited atom, is a fundamental process in quantum electrodynamics (QED), typically associated with exponential decay to the ground state accompanied by irreversible photon emission. This simple Markovian picture, however, is profoundly modified in the presence of time-delayed feedback, structured continua, or cooperative emission, as occurs when an emitter radiates in front of a mirror, when several emitters radiate collectively, or in the case of a giant atom. In such regimes, strong non-Markovian dynamics arise from photon reabsorption and interference effects, leading to pronounced deviations from exponential decay.
Here we demonstrate the emergence of exceptional points (EPs) in these highly non-Markovian waveguide-QED environments, i.e., non-Markovian EPs. These EPs appear directly in the relaxation dynamics as sharp transitions to oscillatory behavior, manifested by the appearance of real zeros in the excited-state amplitude. We analyze in detail the spontaneous emission of giant atoms with two or more coupling points, highlighting the mechanisms leading to non-Markovian EPs, and show that similar phenomena arise in other waveguide-QED settings, such as the collective spontaneous emission of spatially separated point-like emitters. Our results reveal waveguide-QED systems as experimentally accessible platforms for realizing and exploring non-Markovian EP physics.
\end{abstract}

\maketitle

\section{Introduction}

The spontaneous emission of a quantum emitter into a continuum of bosonic modes is a
fundamental process in quantum electrodynamics (QED), typically associated with an
exponential decay of the emitter and irreversible photon emission. This behavior is well
captured by the Weisskopf--Wigner theory~\cite{R1}, which relies on a Markovian
approximation valid for atoms radiating into free space \cite{R1b}. However, this simple
picture is profoundly modified in structured environments or in collective atom
emission, where memory effects, interference, and time-delayed feedback become relevant
\cite{R2,R3,R4,R5,R6,R7,R7b,R8,R9}.

A paradigmatic example is provided by a point-like two-level emitter radiating in front of a
mirror~\cite{R10,R11,R12,R13,R14}. In this case, the finite round-trip time of photons
introduces a time-delayed self-interaction: the emitter effectively interacts with its
own past radiation. This feedback mechanism gives rise to strongly non-Markovian dynamics,
including oscillatory decay, population trapping, and the formation of atom--photon bound
states. Mirror-mediated emission thus provides a minimal and conceptually transparent
platform to investigate how retardation and interference effects reshape spontaneous
emission.

A more general and highly tunable realization of such physics is offered by
\emph{giant atoms}~\cite{G1,G2,G3,G4,G5,G6,G7,G8,G9,G10,G11,G12,G13,G14,G15,G16,G17,G18,G19,G20,
G21,G22,G23,G24,G25,G26,G27,G28,G29,G30,G31,G32,G33,G34,G35,G36,G37,G37b,G37c,G38,G39,G40,G41,G42,
G43,G44,G45,G46,G47,G48,G49}. Giant atoms are artificial quantum emitters -- such as
superconducting transmon qubits \cite{Nori0} -- whose spatial extent exceeds the wavelength
of the emitted radiation, allowing them to couple to a waveguide or transmission line at
multiple, spatially separated points. As a consequence, the emitter experiences non-local
and time-delayed self-interaction even in the absence of mirrors. These delayed couplings
produce strong deviations from exponential decay, including sub- and superradiant
behavior, non-exponential relaxation, and the emergence of atom-photon bound states.
Closely related delayed dynamics also arise in the {cooperative spontaneous emission}
of two or more spatially separated point-like emitters coupled to a common one-dimensional
waveguide~\cite{R7,R7b,R8,R9,Sol1,Sol2,Sol3}. In this case, photons emitted by one atom can be
reabsorbed by another after a finite propagation time, leading to retardation effects,
collective interference, and non-Markovian memory in the joint atomic dynamics. Such
systems provide an alternative and conceptually complementary platform to giant atoms for
exploring time-delayed light-matter interaction in waveguide QED.

Crucially, time-delayed feedback and cooperative emission imply that the dynamics of such
waveguide-QED systems is governed by effective non-Hermitian generators with memory. The
spectral properties of these generators play a central role in determining the
relaxation dynamics. In this context, \emph{exceptional points} (EPs) -- non-Hermitian
degeneracies at which both eigenvalues and eigenvectors of an underlying effective non-Hermiitan Hamiltonian or Liouvillian coalesce~\cite{EP1,EP2,EP3,EP4} -- 
naturally enter the discussion.  EPs and their widespread spectrum of applications -- 
ranging from sensing and metrology to mode-switching, non-reciprocal dynamics, and
topological control -- have been extensively explored in photonics and Markovian open
quantum systems using either effective non-Hermitian or Liouvillian dynamics; see e.g. \cite{EP5,EP6,EP7,EP8,EP9,EP9a,EP9b,EP9c,EP10,EP11,EP12,EP13} and references
therein.
However, their emergence in genuinely non-Markovian quantum dynamics remains
comparatively less explored and has only recently attracted growing interest \cite{NM-3,NM-2,NM-1,NM0,NM1,NM2,NM2b,NM3,NM4,NM5,NM6,NM7,NM8,NM9,NM10}. Here, memory effects invalidate a
time-local description of the dynamics and EPs are no longer associated with finite-dimensional non-Hermitian
Hamiltonians or Liouvillians, but instead emerge from the spectral properties of nonlocal evolution
operators with memory kernels or time delays.
Recent works developed frameworks for identifying EPs in genuinely non-Markovian open quantum systems using pseudomode and hierarchical equations of motion, revealing purely non-Markovian EPs inaccessible in the Markovian limit~\cite{NM3,NM4}. In parallel, waveguide QED platforms have been shown to host strong retardation effects
arising from photon propagation delays and long-range emitter--emitter interactions 
\cite{R10,R11,R12,R13,R14,G1,G2,G3,G4,G5,G6,R7,R7b,R8,R9,Sol1,Sol2,Sol3},
suggesting that they provide a natural arena for the emergence of non-Markovian EPs.
Despite these advances, a clear and physically transparent connection between
retardation-induced memory effects and exceptional-point physics in waveguide QED
remains largely unexplored.

In this work, we demonstrate that non-Markovian exceptional points naturally emerge in
waveguide-QED systems with retardation, focusing on two paradigmatic models. First, we
present a detailed analysis of the spontaneous emission of a giant atom with two or more
coupling points, highlighting the mechanisms leading to non-Markovian dynamics and the
emergence of higher-order exceptional points. We show that these EPs appear at special couplings and critical
delay times, manifesting as sharp transitions to oscillatory relaxation, accompanied by
the appearance of real zeros in the excited-state amplitude. Second, we show that similar
phenomena can occur in the collective spontaneous emission of spatially separated
point-like emitters coupled to a common waveguide.
Although non-Markovian effects in both systems have been previously investigated, the
appearance of non-Markovian EPs has so far been overlooked. Our results establish a direct connection
between non-Markovian spontaneous emission and exceptional-point physics, highlighting
waveguide-QED systems as versatile and experimentally accessible platforms to explore
non-Markovian EP physics.

\section{Non-Markovian exceptional points}
\label{sec:NM_EP_general}

Exceptional points are spectral singularities of non-Hermitian systems at which
both eigenvalues and eigenvectors coalesce, leading to a non-diagonalizable generator
and to characteristic non-exponential dynamics~\cite{EP1,EP2,EP3,EP4}. While EPs are most commonly discussed
in the context of finite-dimensional, time-local non-Hermitian Hamiltonians or
Liouvillian superoperators (see e.g. \cite{EP1,EP2,EP3,EP4,EP5,EP6,EP7,EP8,EP9,EP9a,EP9b,EP9c,EP10,EP11,EP12,EP13} and references therein), analogous phenomena can arise more generally in
\emph{non-Markovian} systems, where the dynamics is intrinsically nonlocal in time \cite{NM-2,NM-1,NM0,NM1,NM2,NM2b,NM3,NM4,NM5}.
{\color{black} Exceptional points in Markovian systems typically arise from the coalescence of eigenvalues and eigenvectors of \emph{finite-dimensional, time-local} non-Hermitian generators, where the dynamics is governed by a finite-order differential equation and the resulting relaxation appears as polynomially modified exponentials. In contrast, non-Markovian EPs occur in systems described by integro-differential equations with memory kernels. Here, the effective generator acts on an \emph{infinite-dimensional history space}, and EPs correspond to the coalescence of poles of the Laplace-domain resolvent. Such EPs are inherently absent in the Markovian limit, because memory or retardation is essential for their formation.} 

A broad class of non-Markovian evolutions can be written in the form of an integro-differential equation
\begin{equation}
\frac{d}{dt}\,\mathbf{x}(t)
=
\int_0^t K(t-t')\,\mathbf{x}(t')\,dt',
\label{eq:memory_kernel_general}
\end{equation}
where $\mathbf{x}(t)$ denotes the system state (e.g., a vector of probability amplitudes
or a vectorized density matrix), and $K(t)$ is a memory kernel encoding retardation,
feedback, or reservoir memory effects. In contrast to Markovian dynamics, the evolution
at time $t$ depends on the entire history of the system over the interval $[0,t]$.
As a consequence, the effective dynamical generator acts on an infinite-dimensional
history space. This time-non-locality complicates the application of traditional spectral analysis techniques
for identifying EPs \cite{NM3}. {\color{black} The ``strength'' of the memory, e.g., the decay time of the kernel $K(t)$ or the delay time in photon-mediated interactions in the models discussed in this work, influences the location of non-Markovian EPs in parameter space, such as the critical delay $\tau_{\rm EP}$ introduced in Sec.III.C, while the order of the EP is determined by symmetry and the number of interaction pathways (Sec.III.E). In spite of this apparent complexity, the dynamics in Eq.~\eqref{eq:memory_kernel_general} can be approximately represented as a \emph{memoryless linear system} in an extended space by introducing auxiliary variables}. For instance, if the kernel can be approximated as a sum of exponentials,
\begin{equation}
K(t) = \sum_{i=1}^N \gamma_i e^{-\lambda_i t},
\end{equation}
one can define auxiliary variables $y_i(t)$ such that
\begin{equation}
\frac{d}{dt} \mathbf{x}(t) = \sum_{i=1}^N y_i(t), \quad
\frac{d}{dt} y_i(t) = -\lambda_i y_i(t) + \gamma_i \mathbf{x}(t)
\end{equation}
($ i=1,\dots,N$).
In this extended $(N+1)$-dimensional system, the dynamics becomes Markovian, with the auxiliary variables encoding the past history of $\mathbf{x}(t)$, and the definition of EPs follows the ordinary route of simultaneous coalesce of two (or more) eigenvalues and corresponding eigenvectors of the linear system Eq.(3). More generally, a continuum of auxiliary variables can be introduced to exactly represent arbitrary kernels, formally mapping the non-Markovian evolution onto a memoryless linear system in an infinite-dimensional space.
The auxiliary variables $y_i(t)$ can be viewed as a mathematical analog of the pseudomodes introduced in Ref.~\cite{NM3}, which also allow a Markovian representation of non-Markovian dynamics. 
Despite this reformulation, exponential functions of the form $\mathbf{x}(t)\propto e^{s t}$ play a
distinguished role. Inserting this ansatz into Eq.~\eqref{eq:memory_kernel_general} yields
an algebraic condition for the complex growth rates $s$, which can be interpreted as
generalized eigenvalues of the non-Markovian evolution.
The structure of the dynamics becomes transparent upon taking the Laplace transform,
\[
\tilde{\mathbf{x}}(s)=\int_0^\infty e^{-s t}\mathbf{x}(t)\,dt .
\]
Equation~\eqref{eq:memory_kernel_general} then leads to
\begin{equation}
\bigl[s\mathbb{I}-\tilde K(s)\bigr]\tilde{\mathbf{x}}(s)=\mathbf{x}(0),
\end{equation}
where $\tilde K(s)$ is the Laplace transform of the memory kernel.
The poles of $\tilde{\mathbf{x}}(s)$, which govern the time evolution via inverse
Laplace transform, are determined by the characteristic equation
\begin{equation}
D(s)\equiv \det\!\bigl[s\mathbb{I}-\tilde K(s)\bigr]=0 .
\label{eq:general_char_eq}
\end{equation}
Each solution $s_n$ corresponds to a dynamical mode contributing a term
$\propto e^{s_n t}$ to the evolution.

In generic situations, the roots $s_n$ of Eq.~\eqref{eq:general_char_eq} are distinct,
and the dynamics is governed by a sum of simple exponentials. A
\emph{non-Markovian exceptional point} arises when two or more of these roots coalesce,
\[
s_1=s_2=\cdots=s_{\mathrm{EP}},
\]
and, simultaneously, the associated dynamical modes become linearly dependent.
In the Laplace-domain formulation, this scenario is signaled by the merging of simple
poles of the resolvent $\bigl[s\mathbb{I}-\tilde K(s)\bigr]^{-1}$ into a higher-order pole.
Equivalently, the condition for the emergence of an exceptional point of order $n \geq 2$ can be expressed as
{\color{black}
\begin{equation}
D(s_{\mathrm{EP}})= D^{\prime}(s_{\mathrm{EP}})= ...= D^{(n-1)}(s_{\mathrm{EP}})=0
\qquad
\label{eq:general_EP_condition}
\end{equation}
where $D^{\prime}(s) \equiv (dD/ds)$ and $D^{(n)}(s) \equiv (d^{n}D/ds^n)$.}
Although the underlying evolution equation may be scalar or low dimensional, the
presence of memory renders the effective spectral problem infinite dimensional.
The coalescing ``eigenvectors'' at a non-Markovian EP are therefore understood as
coalescing \emph{dynamical eigenmodes} in the history space \cite{L2}, rather than as ordinary
state vectors of a finite-dimensional Hilbert space.
The coalescence of poles at a non-Markovian EP has direct consequences in the time
domain. Instead of a sum of purely exponential contributions, the inverse Laplace
transform yields polynomially modified exponentials of the form
\begin{equation}
\mathbf{x}(t)\sim t^{m} e^{s_{\mathrm{EP}} t},
\end{equation}
where $(m-1)$ depends on the order of the EP. Such non-exponential relaxation behavior
constitutes a universal dynamical signature of exceptional points, independent of
whether the system is Markovian or non-Markovian.

Clearly, non-Markovian EPs constitute a generic feature of open quantum systems with memory, retardation, or
coherent feedback. Such conditions naturally arise in waveguide and circuit QED
platforms, where finite propagation times and structured reservoirs lead to
time-delayed interactions and intrinsically non-Markovian dynamics. In the following of this paper, we unveil the occurrence of non-Markovian EPs in a
waveguide-QED setting, focusing on two paradigmatic models. First, we analyze the
spontaneous emission dynamics of a giant atom coupled to a one-dimensional waveguide,
where the spatially separated coupling points give rise to self-interference and
time-delayed feedback. This system provides a natural platform to observe high-order non-Markovian EP under suitable engineering of the 
coupling constants at various contact points.
Second, we briefly comment on other possible QED waveguide settings to observe non-Markovian EPs, such as  the cooperative spontaneous emission of two
spatially separated point-like atoms coupled to a common waveguide, where retardation
and photon-mediated interactions induce collective non-Markovian dynamics.  

\section{Non-Markovian exceptional points in the spontaneous emission of giant atoms}
\subsection{Model}
\label{sec:giant_atom_single_excitation}

Let us consider the relaxation dynamics of a giant atom coupled to a one-dimensional waveguide, a problem discussed in several works (see, e.g., \cite{G1,G4,G37b,G45}) and briefly reviewed here to introduce the model and establish notation.  
We consider a two-level emitter with excited state \(|e\rangle\), ground state \(|g\rangle\), and transition frequency \(\omega_0\), coupled to a one-dimensional waveguide at $N=2$ spatially separated contact points, \(x_1=0\) and \(x_2=d\) [Fig.~1(a)]; the case $N>2$ will be considered in Sec.III.E. The waveguide supports right- and left-propagating modes with linear dispersion \(\omega_k = v_g |k|\), \(-\infty < k <\ \infty \), where \(v_g\) is the group velocity at the atomic transition frequency \(\omega_0\). Within the rotating-wave approximation and setting \(\hbar = 1\), the full atom-photon Hamiltonian reads
\begin{align}
H &= \omega_0 \sigma_+\sigma_- 
+ \int_{-\infty}^{\infty} \mathrm{d}k\, \omega_k\, a_{k}^\dagger a_{ k} \nonumber\\
&\quad + \int_{-\infty}^\infty \mathrm{d}k\, \Big[ g \left(1+e^{ikd} \right)\ \sigma_+  a_{ k} + \text{H.c.} \Big],
\end{align}
where \(\sigma_+ = |e\rangle\langle g| \), \( \sigma_-=|g\rangle\langle e|\) are the atomic raising and lowering operators, \(g\) is the atom-waveguide coupling at each connection point, and  \(a_{k}^\dagger\),  \(a_{k} \) create and annihilate a photon of wavevector \(k \) in the waveguide; they satisfy the canonical bosonic commutation relations \([a_{k}, a_{k'}^\dagger] = \delta(k-k')\).  
Restricting to the single-excitation sector, with the atom initially excited and the waveguide in the vacuum, the system state can be expanded as
\begin{equation}
|\Psi(t)\rangle = a(t) e^{-i \omega_0 t} |e, \mathrm{vac}\rangle +  \int_{-\infty}^\infty \mathrm{d}k\, {\phi}_{k}(t) e^{- i \omega_k t} |g, 1_{k}\rangle,
\end{equation}
where \(a(0)=1\) and \( {\phi}_{ k}(0)=0\).  
The Schr\"odinger equation gives
\begin{align}
\dot{a}(t) &= - i g \int_{-\infty}^\infty \mathrm{d}k \, 
\big(1 + e^{ i k d}\big) {\phi}_{k}(t) 
e^{i (\omega_0 - \omega_k) t}, \\
\dot{{\phi}}_{k}(t) &= - i g \left(1 + e^{- i k d}\right) a(t) e^{- i (\omega_0 - \omega_k) t}.
\end{align}
Formally integrating the field amplitudes gives
\begin{equation}
{\phi}_{k}(t) =- i g (1 + e^{ -i k d}) \int_0^t dt' \, a(t') e^{-i (\omega_0 - \omega_k) t'}.
\end{equation}
Substituting back into the equation for \(a(t)\) yields the integro-differential equation
\[
\dot{a}(t)=\int_{0}^t dt' K(t-t') a(t')
\]
\begin{figure}[t]
 \centering
    \includegraphics[width=0.48\textwidth]{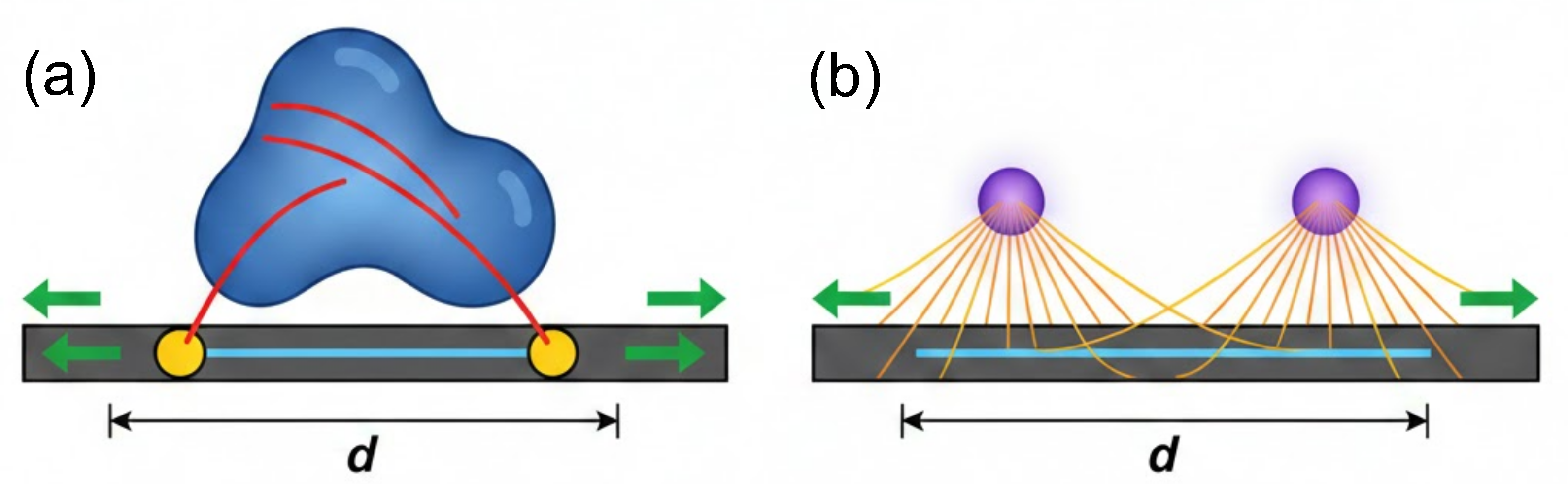}
   \caption{(a) Schematic of a two-level giant atom radiating into a one-dimensional waveguide. The two contact points between the giant atom and the waveguide are spaced by $d$. (b) Schematic of the collective spontaneous emission of two identical point-like two-level emitters into a one-dimensional waveguide. The spacing between the two emitters is $d$.}
    \label{fig1}
\end{figure}
which is of the form (1) with the kernel
\begin{equation}
K(t-t')=-2 g^2 \int_{-\infty}^{\infty} dk \left( 1+\cos kd \right) e^{-i (\omega_0 - \omega_k)(t-t')}.
\end{equation}
Performing the continuum integral over $k$ with the linear dispersion \(\omega_k = v_g |k | \) and disregarding Lamb-shift corrections leads to the closed delay-differential equation:
\begin{equation}
\dot{a}(t) = - \gamma a(t) - \gamma e^{i \varphi} a(t - \tau) \Theta(t - \tau),
\end{equation}
where \(\tau = d / v_g\) is the propagation time delay, \(\gamma /2 = 2 \pi g^2/v_g$ is the single-point atom decay rate, \(\varphi = \omega_0 \tau\) is the phase accumulated during propagation, and \(\Theta(t-\tau)\) is the Heaviside step function enforcing causality [$\Theta(t-\tau)=0$ for $t< \tau$ and $\Theta(t-\tau)=1$ for $t> \tau$].

Equation~(14) is the characteristic differential-delayed equation describing the non-Markovian relaxation dynamics of the giant atom in a rotating frame \cite{G1,G4,G7,G37b}.  The first term corresponds to spontaneous emission at each coupling point, while the second term represents coherent time-delayed self-interaction arising from emission at one point and reabsorption at the other.  The phase factor \( e^{ i \varphi} \) encodes interference between the two emission pathways and plays a central role in determining the decay behavior.
As it is well known,  the dynamics exhibit strong deviations from simple exponential decay and pronounced non-Markovian effects when the time delay \(\tau\) becomes comparable to or larger than the spontaneous lifetime \(1/\gamma\).  Similar non-Markovian behavior arises for point-like emitters placed in front of a mirror.  Furthermore, the delayed feedback induced by the non-local coupling can lead to either sub-radiant or super-radiant spontaneous emission dynamics, and in particular to the formation of atom-photon bound states when the phase delay satisfies \(\varphi = \omega_0 \tau = \pi \, (\mathrm{mod}\, 2\pi)\).

\subsection{Relaxation dynamics}

The solution to the delay-differential equation for the atomic amplitude $a(t)$ in the rotating frame, Eq.(14), with the initial condition $a(0)=1$ can be readily obtained in an exact closed form by recurrence over successive intervals, as shown in some previous works (see e.g. \cite{R7b,R12,G9}).  The solution can be written compactly as
\begin{equation}
a(t) = \sum_{n=0}^{\infty} \frac{\big[-\gamma\, e^{i\varphi} (t-n\tau)\big]^n}{n!}\, e^{-\gamma (t-n\tau)} \, \Theta(t-n\tau),
\end{equation}
where $\Theta(t)$ is the Heaviside step function.
Each term in the series represents the contribution to the atomic amplitude after \(n\) round trips of the emitted photon between the two coupling points, with the phase factor \(e^{i \varphi}\) accounting for the interference accumulated during each delay interval. While this series form clearly highlights the non-Markovian effects arising from interference among different decay pathways, it is not well suited to capture the emergence of non-Markovian exceptional points. An alternative but equivalent representation, which is better suited to reveal EPs, can be obtained using the Laplace transform method, as discussed in Sec.II (see also \cite{R7,R7b,NM5,G45,Sol1,L1,L2}). Applying the Laplace transform,
\[
\tilde{a}(s) = \int_0^\infty a(t) e^{-s t} dt,
\]
to Eq.~(14) then yields
\begin{align}
s \,\tilde{a}(s) - a(0) &= - \gamma \,\tilde{a}(s) - \gamma \, e^{i\varphi} e^{-s \tau} \tilde{a}(s), \nonumber
\end{align}
i.e.,
\begin{equation}
\tilde{a}(s) = \frac{1}{s + \gamma + \gamma \, e^{i\varphi} e^{-s \tau}}.
\end{equation}
The poles of \(\tilde{a}(s)\) satisfy the transcendental equation
\begin{equation}
D(s)=s + \gamma + \gamma e^{i\varphi} e^{-s \tau} = 0.
\label{eq:char_eq}
\end{equation}
The solution to this equation can be written in terms of the 
the Lambert \(W(z)\) function. This is a multi-valued function defined on the complex $z$ plane via the transcendental equation 
\[ W_n(z) e^{W_n(z)} = z , \] 
where 
\(W_n\) denotes the \(n\)-th branch, with $n=0,\pm 1, \pm 2, \pm3...$ (see e.g. \cite{Sol1,L1,L2}).
  \begin{figure*}
 \centering
    \includegraphics[width=0.95\textwidth]{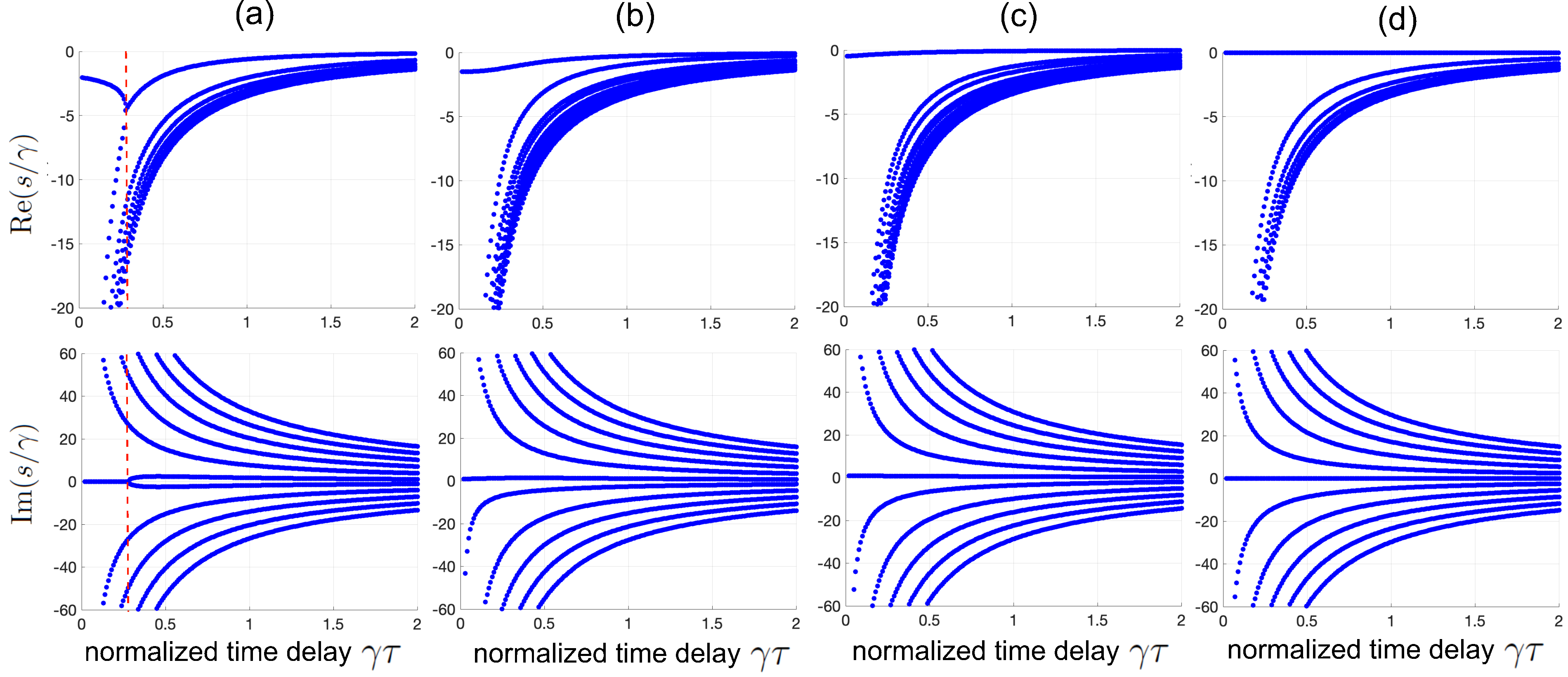}
   \caption{Behavior of a few dominant poles $s=s_n$ (real and imaginary parts) versus the normalized time delay $\gamma \tau$, as given by Eq.(19), for a few increasing values of the phase $\varphi$. (a) $\varphi=0$, (b) $\varphi= \pi/3$, (c) $\varphi= 2 \pi /3$, and (d) $\varphi=\pi$. In the $\varphi=0$ case [panel (a)], a coalescence of the two dominant poles is clearly observed at $\gamma \tau \simeq 0.28$ (vertical red dashed curves), according to the theoretical analysis [Eq.(22)].}
    \label{fig1}
\end{figure*}
Rewriting Eq.(17) in the equivalent form
\begin{equation}
(s + \gamma) \tau e^{(s+\gamma)\tau} = - \gamma \tau e^{i\varphi} e^{\gamma \tau}.
\end{equation}
one then obtains an infinite number of poles at $s=s_n$, with ${\rm Re}(s_n) \leq 0$, where
\begin{equation}
s_n = -\gamma + \frac{1}{\tau} W_n \Big( - \gamma \tau e^{\gamma \tau} e^{i\varphi} \Big)
\end{equation}
 and $n=0, \pm 1, \pm 2, \pm 3...$.  Each branch corresponds to a different mode contributing to the non-Markovian dynamics.
 The time-domain solution is obtained by performing the inverse Laplace transform of
$\tilde a(s)$, which can be evaluated by closing the Bromwich contour in the complex
$s$ plane and summing the residues at the poles $s=s_n$. Assuming simple poles, one  then finds
\begin{equation}
a(t)
=
\sum_{n=-\infty}^{\infty}
\frac{e^{s_n t}}{1-\gamma \tau e^{i\varphi} e^{-s_n \tau}},
\label{eq:a_t_laplace_solution}
\end{equation}
where the denominator arises from the derivative of the characteristic function in
Eq.~(16) evaluated at $s=s_n$. Each term in the sum represents an
eigenmode of the non-Markovian dynamics associated with a distinct branch $W_n$ of
the Lambert function. The long-time relaxation is governed by the poles with the
largest real part, while the full sum reproduces the exact non-Markovian evolution
at all times. Equation~\eqref{eq:a_t_laplace_solution} is fully equivalent to the series solution
in Eq.~(15). While Eq.~(15) makes explicit the time-domain interpretation in terms of
successive delayed emission processes, Eq.~\eqref{eq:a_t_laplace_solution} provides a
spectral decomposition of the dynamics in terms of the eigenmodes (pseudo-modes) $s_n$ of the
non-Markovian evolution. The equivalence of the two forms \cite{R7,R7b} follows from the fact that
both originate from the same delay-differential equation and represent different
resummations of the same physical processes: interference between instantaneous and
time-delayed decay channels.

\subsection{Non-Markovian exceptional points}

The expression in Eq.~\eqref{eq:a_t_laplace_solution} assumes that all poles $s_n$ of
$\tilde a(s)$ are simple. In this case, each pole contributes a purely exponential
term to the dynamics. When two (or more) poles coalesce, however, the residue theorem
must be generalized to account for higher-order poles. As a result, the time-domain
solution acquires polynomially modified exponentials of the form
$t^m e^{s_{\mathrm{EP}} t}$, where $s_{\mathrm{EP}}$ denotes the degenerate eigenvalue
and $(m-1)$ is determined by the order of the exceptional point. Such non-exponential
relaxation is a characteristic signature of exceptional-point dynamics.
 As discussed in Sec.II, while the evolution equation for $a(t)$ is scalar, the presence of a finite time
delay renders the dynamics effectively infinite dimensional, since the state at time
$t$ depends on the full history of the system over the interval $[t-\tau,t]$.
Exponential solutions of the form $a(t)\propto e^{s t}$ therefore play the role of
eigenmodes of the non-Markovian evolution operator acting on this history space \cite{L2}, with
the corresponding eigenvalues given by the roots $s_n$ of the characteristic
equation~\eqref{eq:char_eq}. Distinct roots correspond to linearly independent
dynamical modes. From this spectral perspective, a non-Markovian exceptional point occurs when two solutions of
Eq.~\eqref{eq:char_eq} coalesce and, at the same time, their associated eigenmodes
become linearly dependent. In the Laplace-domain representation, this scenario is
signaled by the merging of two simple poles of $\tilde a(s)$ into a higher-order pole.
The EP condition is therefore given by
\begin{equation}
D(s)= s + \gamma + \gamma e^{i\varphi} e^{-s \tau} = 0,
\qquad
\frac{dD}{ds}=0,
\label{eq:EP_condition}
\end{equation}
which ensures both eigenvalue degeneracy and modal coalescence. This yields the trascendental equation
\[
\gamma \tau e^{1+ \gamma \tau+ i \varphi}=1
\]
which is satisfied provided that $\varphi=0$ (mod $ 2 \pi$), i.e. in the so-called super-radiant emission regime, and \(
W_0( \gamma \tau)=1/e\),
i.e. 
\begin{equation}
\tau= \tau_{\mathrm{EP}} \approx 0.27846/\gamma.
\end{equation}
Figure 2 shows the behavior of the poles $s_n$ versus $ \gamma \tau$ for a few values of the retardation phase $\varphi$. As one can see, a coalescence of the first two dominant poles, corresponding to the crossing of the branches $W_0$ and $W_{-1}$ of the Lambert function in Eq.(19), occurs at $\tau=\tau_{\mathrm{EP}}$, as predicted by Eq.(22). Hence a {\color{black} second-order} non-Markovian EP arises in the super-radiant regime $\varphi=0$ when the delay time $\tau$ is tuned at the value $\tau_{\mathrm{EP}}$ given by Eq.(22). 

\begin{figure}
 \centering
    \includegraphics[width=0.48\textwidth]{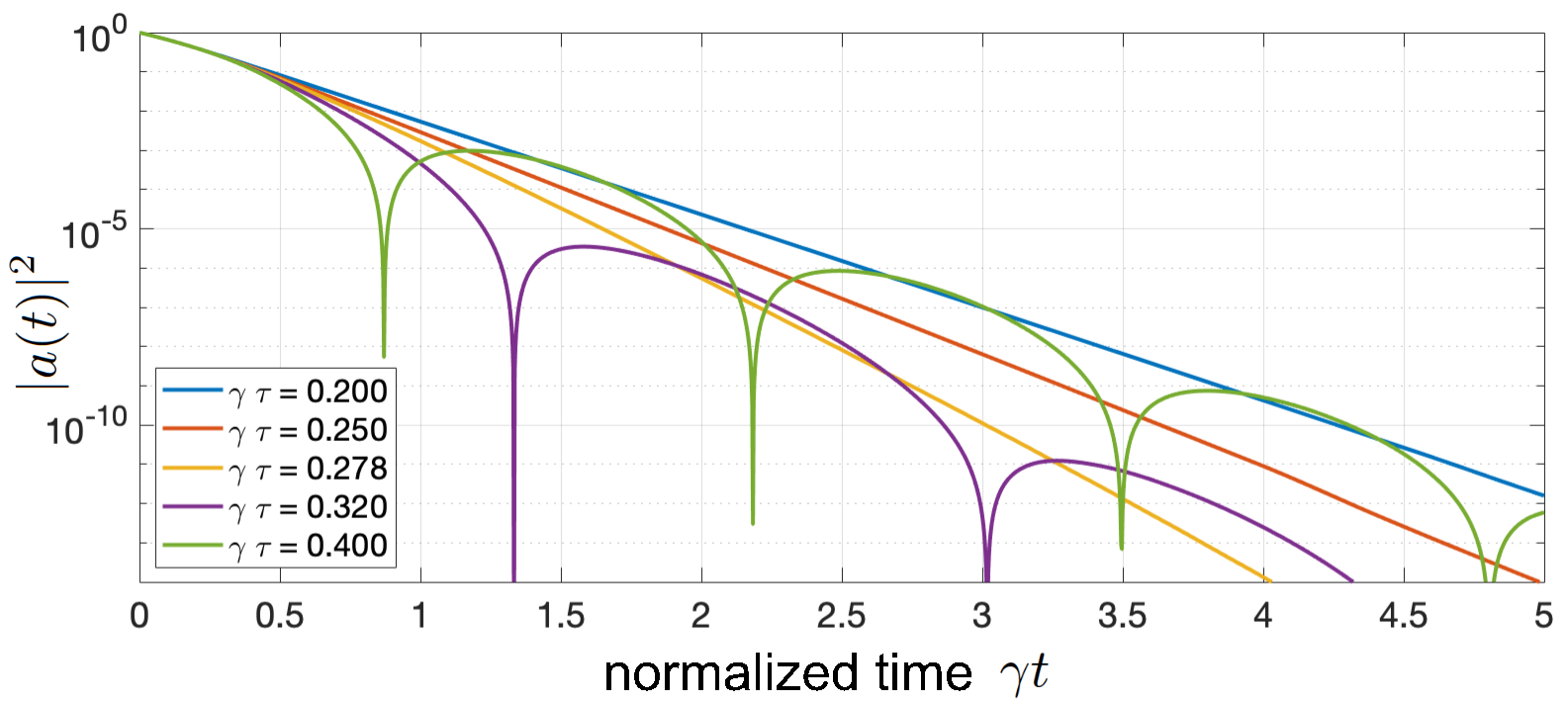}
   \caption{Numerically-computed spontaneous emission decay of the giant atom (behavior of $|a(t)|^2$ on a vertical log scale versus normalized time $\gamma t$) for $\varphi=0$ and for a few increasing values of the normalized time delay $\gamma \tau$. As the EP value $\gamma \tau_{\mathrm{EP}} \simeq 0.27846$ is crossed, a transition to oscillatory damped oscillations is clearly visible.}
    \label{fig3}
\end{figure}

The coalescence of the two dominant poles at $\tau=\tau_{\mathrm{EP}}$ has a direct and
experimentally observable manifestation in the time-domain dynamics of the excited-state
amplitude $a(t)$. For $\tau<\tau_{\mathrm{EP}}$, the two leading poles are real and negative,
and the relaxation dynamics is purely monotonic, with $a(t)$ decaying without zeros.
For $\tau>\tau_{\mathrm{EP}}$, the poles form a complex-conjugate pair $s_{\pm} = -\Gamma \pm i\Omega$,
leading to oscillatory relaxation and the appearance of multiple real zeros of $a(t)$
due to interference between the instantaneous decay and the delayed emission pathway;
see Fig.~3 for illustration. Close to the exceptional point, the imaginary part of the
dominant poles exhibits the characteristic square-root scaling
\begin{equation}
\Omega \propto \sqrt{\tau-\tau_{\mathrm{EP}}},
\end{equation}
{\color{black} which is typical of a second-order EP in Markovian systems}.
Correspondingly, the oscillation period diverges as
\begin{equation}
T = \frac{2\pi}{\Omega} \sim \frac{1}{\sqrt{\tau-\tau_{\mathrm{EP}}}}
\qquad (\tau \to \tau_{\mathrm{EP}}^{+}).
\end{equation}
At the EP itself, oscillations disappear and the dynamics is governed by a polynomially
modified exponential, $a(t) \sim t\, e^{s_{\mathrm{EP}} t}$, reflecting the presence of a
second-order pole in the Laplace domain. This critical slowing down is a universal
temporal signature of exceptional points and is independent of the specific realization
of the delay. We note that such enhanced dynamical sensitivity near EPs may offer
opportunities for parameter estimation or sensing in waveguide-QED platforms
\cite{sensing1,sensing2}. {\color{black} Finally, it is worth noting that in such a waveguide-QED system the interplay between retardation and interference effects can give rise to two qualitatively distinct types of oscillatory dynamics. The first type consists of oscillations induced by atom-photon bound states, corresponding to isolated poles on the imaginary axis, which produce undamped dynamics. The second type consists of oscillatory relaxation triggered by the coalescence of poles at an exceptional point (EP) discussed above, arising from interference and retardation effects. These mechanisms are physically distinct and do not coexist: non-Markovian EPs emerge in the super-radiant phase ($ \varphi= 0$), whereas atom-photon bound states appear in the sub-radiant phase ($ \varphi= \pi$, corresponding to the $s=0$ pole), so that their coexistence is prevented in the models considered here.}

\subsection{Pseudomode interpretation}

To gain further physical insight into the exceptional point identified above, it is instructive to formulate an approximate, memoryless model that captures the dominant spectral features of the non-Markovian evolution via the pseudo-mode approach discussed in Sec.II. The Laplace transform $\tilde{a}(s)$ of the atomic amplitude
has an infinite set of poles \(s_n\) determined by Eq.~(19). As discussed above, the dynamics is mainly governed by the two poles with the largest real parts, corresponding to the \(n=0\) and \(n=-1\) branches of the Lambert function. In the vicinity of the exceptional delay \(\tau_E\), these two poles approach one another and coalesce at the EP.
In a {\em two-pole approximation}, one retains only these dominant spectral contributions and approximates \(\tilde a(s)\) by a rational function with two simple poles,
\begin{equation}
\tilde{a}(s)\approx\frac{A}{(s-s_1)(s-s_2)}\,,
\end{equation}
where \(s_1,s_2\) are the two leading solutions of Eq.~(19), and the prefactor \(A\) is fixed by matching residues at those poles. Inverse Laplace transforming this expression yields
\[
a(t)\approx\frac{A}{s_1-s_2}\,\big(e^{s_2 t}-e^{s_1 t}\big),
\]
which satisfies the {\em memoryless} second-order differential equation
\begin{equation}
\ddot a(t)-(s_1+s_2)\dot a(t)+s_1s_2\,a(t)=0\,.
\end{equation}
Introducing the auxiliary variable \(b(t)=\dot a(t)-s_1 a(t)\), or equivalently the vector
\[
\mathbf{x}(t)=\begin{pmatrix}a(t)\\[1ex]b(t)\end{pmatrix},
\]
brings this into a {\em first-order Markovian system}
\begin{equation}
\dot{\mathbf{x}}(t)=
\begin{pmatrix}
0 & 1\\[1ex]
-s_1s_2 & s_1+s_2
\end{pmatrix}
\mathbf{x}(t)\,.
\label{eq:two_mode_system}
\end{equation}
The two eigenvalues of the system matrix are precisely \(s_1,s_2\), and \(b(t)\) can be interpreted as an effective {\em pseudomode} that encodes the leading memory effects of the original delay dynamics. This construction is analogous to generalized pseudomode mappings in open quantum systems, where the environment's influence is distilled into a small set of discrete modes whose dynamics is Markovian on an enlarged space \cite{NM3,NM4}. Within this reduced description, the non-Markovian EP appears as an ordinary second-order EP of the \(2\times 2\) system in Eq.~\eqref{eq:two_mode_system}. The EP condition \(s_1=s_2\) renders the matrix non-diagonalizable with a single Jordan block, and the time evolution then contains a polynomially modified exponential, \(
a(t)\sim t\,e^{s_{\mathrm{EP}}t}\,,\)
in agreement with the inverse Laplace analysis of the full delay equation at coalescence (Sec.~III.C). Thus, the two-pole approximation provides a concrete, memoryless linear system that reproduces both the qualitative dynamics and the exceptional-point behavior of the original non-Markovian model, and shows how pseudomode concepts can connect the infinite-dimensional delay formulation to a finite-dimensional Markovian picture.

\subsection{Extension to \(N\) contact points and higher-order non-Markovian exceptional points}

The delay-differential equation (14) can be extended to the case of a giant atom coupled to a one-dimensional waveguide at \(N\) spatially separated points.  Labeling the contact points by \(j=1,\dots,N\) at positions
\(
x_j=(j-1)\,d,
\)
with fixed spacing \(d\), and allowing for possibly distinct real coupling strengths \(g_j\) at each point, the atomic amplitude \(a(t)\) in the single-excitation sector satisfies a multi-delay differential equation incorporating all propagation pathways between the contact points \cite{G10,G49}.  Defining
\(\tau=d/v_g$ and \( \varphi\equiv\omega_0\tau,\)
one finds (see e.g. \cite{G49})
\begin{equation}
\dot a(t)
=-\sum_{j,\ell=1}^{N}
\Gamma_{j\ell}\,
e^{\,i\varphi\,|j-\ell|}\,
a\bigl(t-|j-\ell|\tau\bigr)\,
\Theta\bigl(t-|j-\ell| \tau\bigr),
\label{eq:N_delay_general}
\end{equation}
with
{\color{black}
\[
\Gamma_{j\ell}\equiv\frac{\pi\,g_j\,g_\ell}{v_g},
\]
}
and the Heaviside step function \(\Theta(\cdot)\) enforcing causality.  Physically, emission at point \(j\) propagates to point \(\ell\) with delay \(|j-\ell|\tau\) and accumulates a phase \(e^{i\varphi|j-\ell|}\) due to propagation between the points.  Grouping terms by delay classes \(m=|j-\ell|=0,1,\dots,N-1\) gives the equivalent representation
\begin{equation}
\dot a(t)
=-\sum_{m=0}^{N-1}\kappa_m\,e^{\,i\varphi m}\,
a(t-m\tau)\,\Theta(t-m\tau),
\label{eq:N_delay_compact}
\end{equation}
with
\begin{equation}
\kappa_m
\equiv
\sum_{\substack{j,\ell=1\\|j-\ell|=m}}^N
\frac{\pi\,g_j\,g_\ell}{v_g}.
\end{equation}
This relation 
 reduces to the familiar \(N=2\) form, considered in the previous subsections, when \(g_1=g_2=g\) real.  Equation~\eqref{eq:N_delay_compact} is a scalar delay equation containing up to \((N-1)\) discrete delays \(|j-\ell|\tau\), giving rise to a rich non-Markovian spectrum.
Taking the Laplace transform \(\tilde a(s)=\int_0^\infty e^{-st}a(t)\,dt\) of Eq.~\eqref{eq:N_delay_compact} yields
\begin{equation}
s\,\tilde a(s)-a(0)
=-\sum_{m=0}^{N-1}\kappa_m\,e^{\,i\varphi m}\,
e^{-s m\tau}\,\tilde a(s),
\end{equation}
so that
\begin{equation}
\tilde a(s)=
\frac{1}{s+\tilde K(s)},
\label{eq:tilde_a_N}
\end{equation}
where the Laplace kernel entering the characteristic equation is
\begin{equation}
\tilde K(s)
\equiv 
\sum_{m=0}^{N-1}\kappa_m\,e^{\,i\varphi m}\,
e^{-s m\tau}.
\label{eq:general_K_s}
\end{equation}
The poles of \(\tilde a(s)\), which satisfy
\begin{equation}
D(s)\equiv s+\tilde K(s)=0,
\label{eq:general_char_N}
\end{equation}
determine the non-Markovian spectrum and the time evolution via the inverse Laplace transform.  Equation~\eqref{eq:general_K_s} generalizes the \(N=2\) kernel \(s+\gamma+\gamma e^{i\varphi}e^{-s\tau}\) appearing in Eq.~(16) of Sec.~III.B.  The multi-exponential structure of \(\tilde K(s)\) reflects the several discrete delayed feedback pathways and the associated interference effects.
Distinct roots of Eq.~\eqref{eq:general_char_N} correspond to independent exponential modes of the non-Markovian evolution.  As in the \(N=2\) case, a non-Markovian EP occurs when two or more of these roots coalesce and the corresponding dynamical modes become linearly dependent (Sec.~III.C).  For \(N>2\), the increased number of discrete delays enables the coalescence of three or more poles of \(\tilde a(s)\), giving rise to higher-order non-Markovian EP. As shown in the Appendix A, rather generally one can engineer the couplings $g_j$ such that an $N$-order non-Markovian EP arises at some target real pole $s=s_{\mathrm{EP}}<0$ and sufficiently long delay time $\tau=\tau_{\mathrm{EP}}$ by assuming $\varphi=0$. Such higher-order EP can also be interpreted within an approximate pseudomode picture by retaining multiple dominant poles of \(\tilde a(s)\) and constructing an equivalent finite-dimensional, Markovian system with multiple coupled modes, in direct analogy to the two-pole pseudomode model discussed in Sec.III.D.

As an illustrative example, let us consider the $N=3$ case. The Laplace-domain characteristic equation determining the non-Markovian spectrum reads explicitly
\[
D(s)=s+\tilde K(s)=s+\kappa_0+\kappa_1e^{i\varphi}e^{-s\tau}+\kappa_2e^{2i\varphi}e^{-2s\tau} =0.
\]
where $\kappa_m$ are the delay weights defined by Eq.(30). A third-order non-Markovian EP corresponds to the coalescence of three characteristic roots of $D(s)$, and is obtained by letting \[
D(s_{\rm EP})=D'(s_{\rm EP})=D''(s_{\rm EP})=0.\]
Such conditions can be satisfied rather generally using the reverse-engineering procedure described in the Appendix A, or by directly imposing their satisfaction via solvability conditions. Here we use the latter approach, which is rather simple for the $N=3$ case.  Specifically, it can be readily shown that a third-order EP occurs at 
\begin{figure}
 \centering
    \includegraphics[width=0.48\textwidth]{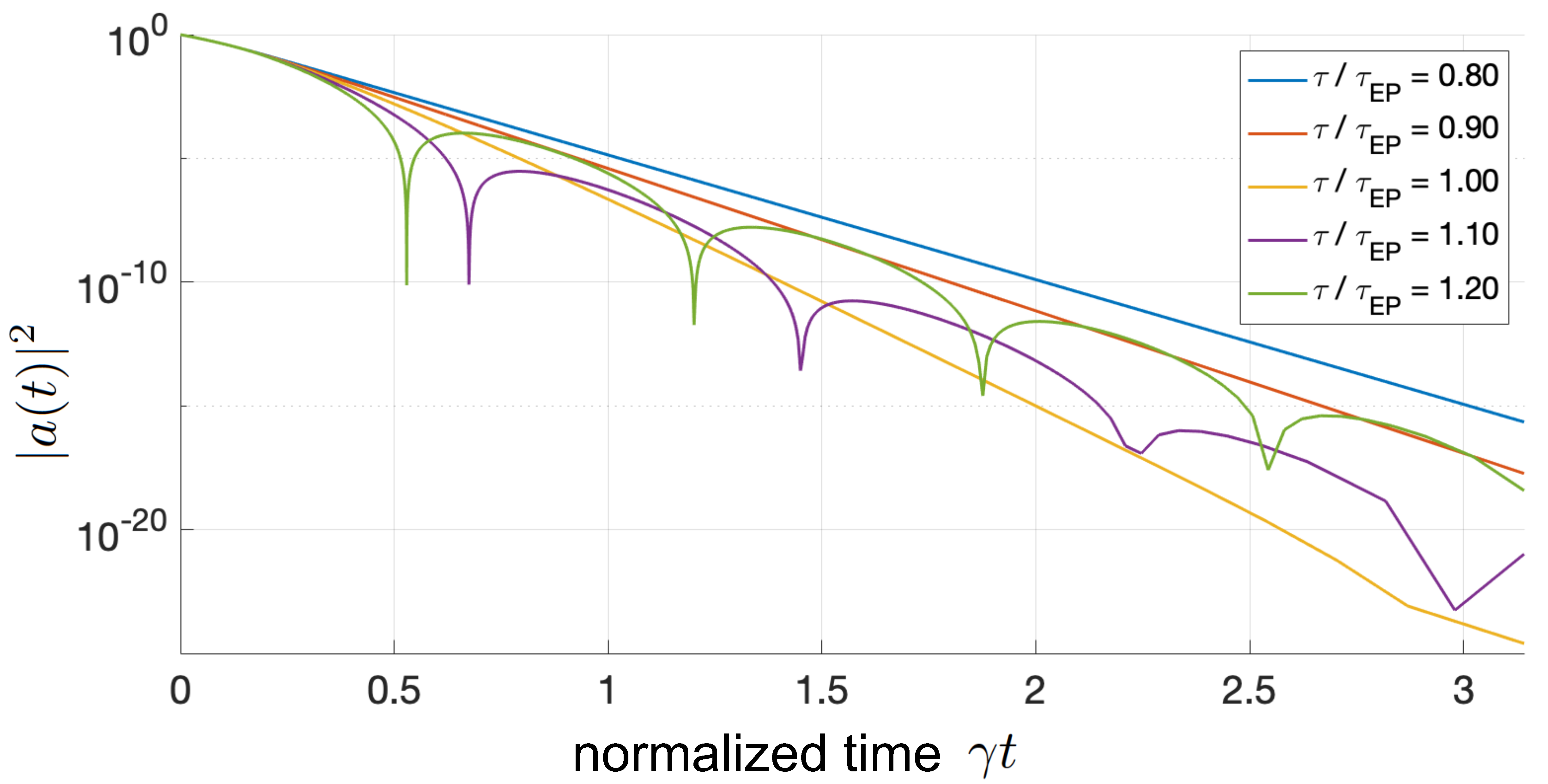}
   \caption{Numerically-computed spontaneous emission decay of a giant atom with three contact points for $g_2/g_1=1$, $g_3/g_1=-0.03846 $, $\varphi=0$ and for a few increasing values of the time delay $\tau$. A third-order non-Markovian EP occurs at the time delay $\tau=\tau_{\mathrm{EP}} \simeq 0.1665 / \gamma$, where we have set $\gamma= \pi g_1^2/v_g$.}
    \label{fig4}
\end{figure}
$s=s_{\mathrm{EP}}$, with
\begin{equation}
s_{\mathrm{EP}}=-\kappa_0-\frac{3}{2 \tau},
\end{equation}
provided that the two solvability conditions
\begin{eqnarray}
\frac{\tau \kappa_1^2}{8 \kappa_2} &= & -1 \\
e^{\kappa_0 \tau+ i \varphi+3/2}  & = &  - \frac{\kappa_1}{4 \kappa_2}
\end{eqnarray}
are satisfied, where
\begin{eqnarray}
\kappa_0 & = & \frac{ \pi}{v_g} (g_1^2+g_2^2+g_3^2)  \\
\kappa_1 & = & \frac{2 \pi}{v_g} (g_1g_2+g_2g_3)  \\
\kappa_2 & = & \frac{2 \pi}{v_g} g_1g_3. 
\end{eqnarray}
Substitution of Eqs.(38-40) into Eqs.(36) and (37) yields the following final form for the solvability conditions
\begin{eqnarray}
e^{- \frac{4g_1g_3(g_1^2+g_2^2+g_3^3)}{g_2^2(g_1+g_3)^2}+\frac{3}{2}+i \varphi} = -\frac{g_2(g_1+g_3)}{4g_1g_3} \\
\tau=- \frac{4v_g}{\pi} \frac{g_1g_3}{g_2^2(g_1+g_3)^2}.
\end{eqnarray}
By letting $\varphi=0$, condition (41) provides a constraint on the coupling constants $g_1,g_2,g_3$, whereas Eq.(42) yields the condition on the delay time $\tau=d/v_g$. Note that, since $\tau$ should be positive, the condition $g_1g_3<0$ is necessary to obtain a third-order EP. For example, assuming $g_1=g_2$ and $\varphi=0$, the third-order EP is obtained by tuning the coupling $g_3$ and the delay time $\tau$ to the values $g_3 \simeq -0.03846 g_1$ and $\tau=\tau_{\mathrm{EP}} \simeq 0.1665 / \gamma$, where we have set $\gamma= \pi g_1^2/v_g$. As an example, Fig.4 shows the numerically-computed spontaneous emission decay in a giant atom with three contact points for $g_2=g_1$, $g_3=-0.03846 g_1$, $\varphi=0$ and for a few increasing values of the time delay $\tau$ crossing the third-order EP point value $\tau_{\mathrm{EP}}=\simeq 0.1665 / \gamma$. As one can see, crossing the EP corresponds to a transition to oscillatory relaxation, with the fastest decay rate obtained at the EP.

\section{Non-Markovian exceptional points in collective spontaneous emission}
 A second paradigmatic example displaying non-Markovian EPs is provided by the collective
spontaneous emission of two spatially-separated point-like emitters, with the same transition frequency $ \omega_0$, coupled to a
common one-dimensional waveguide; see Fig.1(b) for a schematic. This model has been introduced and deeply investigated in some recent works ~\cite{Sol1,Sol3}, however the occurrence of EPs was not highlighted.

Let $c_{1,2}(t)$ denote the probability amplitudes for the two atoms to be in their
excited states, assuming the waveguide initially in the vacuum. In the single-excitation sector and after tracing out
the photonic degrees of freedom, the amplitudes obey the delay-differential equations
\cite{Sol1,Sol2,Sol3}
\begin{eqnarray}
\dot c_1(t)
& = &  -\frac{\gamma}{2}
\left[
c_1(t)
+ \beta\, c_2 \!\left(t-\tau \right)
e^{i\phi_p}\,
\Theta\!\left(t- \tau \right)
\right] \\ 
\dot c_2(t)
& = &  -\frac{\gamma}{2}
\left[
c_2(t)
+ \beta\, c_1 \!\left(t-\tau \right)
e^{i\phi_p}\,
\Theta\!\left(t- \tau \right)
\right],
\end{eqnarray}
 where $\gamma$ is the single-atom decay rate into the waveguide, $ \tau= d/v_g$ is the time delay, $v_g$ is the group
velocity,  $d$ is the spatial separation between the two atoms, and $\Theta(t)$ is the Heaviside step function. The phase
$\phi_p=\omega_0 \tau$ accounts for propagation between the atoms, while the
dimensionless parameter $0\le\beta\le1$ quantifies the fraction of spontaneous
emission mediated by the waveguide ($\beta=1$ corresponds to ideal coupling).
Following Ref.~\cite{Sol1}, it is convenient to introduce the dimensionless distance
parameter
\begin{equation}
\eta = \frac{\gamma d}{ v_g}
= \gamma \tau,
\end{equation}
which measures the propagation delay $\tau$ in units of the spontaneous lifetime $ 1 / \gamma$. 
For the super-radiant initial condition $c_1(0)=c_2(0)=1/ \sqrt{2}$ and $\phi_p=0$, symmetry ensures that
$c_1(t)=c_2(t)\equiv a(t)$ at all times. Equations~(43,44)  then
reduce to a single delay-differential equation,
\begin{equation}
\dot a(t)
= -\frac{\gamma}{2}\, a(t)
- \frac{\gamma}{2}\,\beta\, e^{i\phi_p}\,
a(t-\tau)\, \Theta(t-\tau),
\label{eq:collective_delay}
\end{equation}
which is formally identical to the spontaneous-emission equation of a giant atom with
two coupling points, with an effective feedback strength proportional to $\beta$.
The ideal case $\beta=1$ maps exactly onto the giant-atom model discussed in Sec.III [Eq.(14)].
Applying the Laplace transform to Eq.~\eqref{eq:collective_delay} yields
\begin{equation}
\tilde a(s)
= \frac{1}{\sqrt{2}} \left( \frac{1}{s + \frac{\gamma}{2}
+ \frac{\gamma}{2}\beta\, e^{i\phi_p} e^{-s\tau}} \right),
\end{equation}
whose poles are determined by the characteristic equation
\begin{equation}
D(s)
= s + \frac{\gamma}{2}
+ \frac{\gamma}{2}\beta\, e^{i\phi_p} e^{-s\tau}
=0.
\label{eq:two_atom_char}
\end{equation}
The transcendental equation (48) can be solved explicitly in terms
of the Lambert \(W\) function. Rewriting it as
\[
\left(s+\frac{\gamma}{2}\right)\tau
\exp\!\left[\left(s+\frac{\gamma}{2}\right)\tau\right]
=
-\frac{\gamma\tau}{2}\,
\beta\, e^{\gamma\tau/2+i\phi_p},
\]
one obtains an infinite set of solutions
\begin{equation}
s_n
=
-\frac{\gamma}{2}
+
\frac{1}{\tau}
W_n\!\left(
-\frac{\gamma\tau}{2}\,
\beta\, e^{\gamma\tau/2+i\phi_p}
\right)
\label{eq:two_atom_poles}
\end{equation}
($n=0,\pm1,\pm2,\ldots ,
$), where \(W_n(z)\) denotes the \(n\)-th branch of the Lambert function. Each branch
corresponds to a distinct dynamical mode of the non-Markovian evolution. \\
In the super-radiant regime, \(\phi_p=0\), the two dominant poles originate from the
principal branches \(W_0\) and \(W_{-1}\). As the dimensionless distance
\(\eta=d \gamma / v_g \) between the two atoms is increased, these two poles approach each other and
coalesce at the critical distance \(\eta=\eta_c\), where the argument of the Lambert
function reaches the branch-point value \(-1/e\). The critical value $\eta_c$ is thus found from the condition
\[
\frac{1}{2}\beta\, \eta_c\, e^{\eta_c/2} = \frac{1}{e},
\] 
i.e.
\begin{equation}
\eta_c=2 W_0 \left(  \frac{1}{e \beta} \right)
\end{equation}
which coincides with the critical inter-emitter separation identified in
Ref.~\cite{Sol1}. 
 We mention that Eq.(50) can be also readily obtained by letting $D(s)=D'(s)=0$. 
This pole coalescence marks the
non-Markovian exceptional point and corresponds to the largest instantaneous decay rate of the super-radiant state, as discussed earlier in Ref.~\cite{Sol1}. 

\section{Conclusion}
In this work, we have shown that non-Markovian exceptional points naturally arise in
waveguide-QED systems with time-delayed feedback or collective emission. By analyzing
the spontaneous emission of a giant atom and the cooperative emission of two spatially
separated emitters, we have demonstrated that these EPs manifest as sharp transitions in
the relaxation dynamics, including the emergence of oscillatory behavior and real zeros
in the excited-state amplitude. While non-Markovian effects due to retardation,
feedback, and photon reabsorption have been extensively investigated in the literature,
the appearance of exceptional points in this context has, to the best of our knowledge,
been largely overlooked. Our results show that retardation-induced memory effects in
waveguide-QED systems naturally give rise to non-Hermitian spectral degeneracies, with
non-Markovian EPs  -- also of high order-- directly encoded in the emitter dynamics.

The findings reported here suggest several avenues for future investigation. Non-Markovian
EPs could be exploited for enhanced control of light-matter interactions, including
tailored decay dynamics, robust state preparation, or sensing strategies in structured
photonic environments. Extending the analysis to multiple emitters, more complex
waveguide geometries, or driven-dissipative regimes may uncover richer manifestations
of non-Markovian EP physics and broaden the range of potential applications in quantum
technologies. More generally, our work highlights waveguide QED as a versatile platform
for exploring the interplay of memory, coherence, and non-Hermitian physics in open
quantum systems.\\
\\
\noindent
\textbf{Acknowledgments}\\
The author acknowledges the
Spanish State Research Agency, through the Severo
Ochoa and Maria de Maeztu Program for Centers and
Units of Excellence in R\&D (Grant No. MDM-2017-0711).\\
\\
\noindent
\textbf{Conflict of Interest}\\
The author declares no conflict of interest.\\
\\
\noindent
\textbf{Data Availability Statement}\\
The data that support the findings of this study are available from the corresponding author upon reasonable request.

\appendix
\section{General construction of $N$-th order non-Markovian exceptional points for a giant atom}

In this Appendix, we show that a giant atom with $N$ spatially separated coupling points can be engineered to exhibit an $N$-th order non-Markovian EP at a preassigned real pole $s=s_{\rm EP}<0$ for a suitable time delay $\tau=\tau_E$. We consider a giant atom coupled to a one-dimensional waveguide at $N$ points, with individual coupling strengths $g_j$ ($j=1,\dots,N$) and a uniform propagation delay $\tau$ between consecutive points. The characteristic equation for the system poles reads $D(s)=0$, where

\begin{equation}
D(s) \equiv s + \tilde K(s), \qquad 
\tilde K(s) = \sum_{m=0}^{N-1} \kappa_m\, e^{im \varphi} e^{-s m \tau}.
\label{eq:appendix_D_s}
\end{equation}
 The coefficients $\kappa_m$ encode the collective coupling to the waveguide and are defined as
\[
\kappa_m \equiv \sum_{\substack{j,\ell=1 \\ |j-\ell|=m}}^N \frac{\pi}{v_g} g_j g_\ell,
\]
i.e.
 \begin{align}
\kappa_0 &= \frac{\pi}{v_g} \sum_{j=1}^{N} g_j^2 , \\
\kappa_m &
= \frac{2\pi}{v_g} \, \sum_{j=1}^{N-m} g_j g_{j+m} \;\; (m=1,2,...,N-1)
\end{align}
where $v_g$ is the group velocity in the waveguide. \\  
To engineer an $N$-th order EP at $s = s_{\rm EP}$, we impose the standard degeneracy conditions
\begin{equation}
D(s_{\rm EP}) = D'(s_{\rm EP}) = \dots = D^{(N-1)}(s_{\rm EP}) = 0.
\label{eq:appendix_EP_conditions}
\end{equation}
Explicitly, these yield a linear Vandermonde-like system of $N$ equations for the unknown coefficients $\kappa_0,\dots,\kappa_{N-1}$:
\begin{equation}
\begin{aligned}
D(s_{\rm EP}) &= s_{\rm EP} + \sum_{m=0}^{N-1} \kappa_m e^{i m \varphi} e^{-s_{\rm EP} m \tau} = 0,\\
D'(s_{\rm EP}) &= 1 - \sum_{m=1}^{N-1} m \tau \, \kappa_m e^{i m \varphi} e^{-s_{\rm EP} m \tau} = 0,\\
D''(s_{\rm EP}) &= \sum_{m=1}^{N-1} (m \tau)^2 \, \kappa_m e^{i m \varphi} e^{-s_{\rm EP} m \tau} = 0,\\
&\;\;\vdots\\
D^{(N-1)}(s_{\rm EP}) &= \sum_{m=1}^{N-1} (m \tau)^{N-1} \, \kappa_m e^{i m \varphi} e^{-s_{\rm EP} m \tau} = 0.
\end{aligned}
\label{eq:appendix_linear_system}
\end{equation}
Since this system is linear in the $\kappa_m$, it can be solved straightforwardly by the Gauss method to obtain the effective couplings $\kappa_0$, $\kappa_1$,..., $\kappa_{N-1}$ that realize the desired EP. Real values of $\kappa_m$ are obtained  by letting $\varphi=0$. Once the coefficients $\kappa_m$ are determined, the physical couplings $g_j$ at the $N$ contact points can be reconstructed by inverting Eqs.(A2,A3), which correspond to a quadratic Hankel-type system. Since this system is quadratic, it can admit zero, one or more physically acceptable solutions ($g_j$ real). A main result that can be readily proven is that, provided that the time delay $\tau=\tau_{\mathrm{EP}}$ is chosen long enough, the Hankel system surely admits of at least one acceptable (physical) solution.
In fact, provided that the delay $\tau=\tau_{\mathrm{EP}}$ is chosen sufficiently large that $\tau |s_{\mathrm{EP}}| \gg 1$, the solution to the linear system Eq.(A5) scales as $\kappa_{m+1}/ \kappa_m \sim \exp(s_{\mathrm{EP}} \tau) \ll 1$, i.e. $| \kappa_0| \gg |\kappa_1| \gg ...\gg |\kappa_{N-1}|$. This follows directly from the structure of the EP linear system when $|s_{\mathrm{EP}}| \tau \gg 1$. In this case, Eqs.(A2) and (A3) can be always inverted with the approximate solution
\[
g_1 \simeq \sqrt{ \frac{\kappa_0 v_g}{ \pi}} \;, \;\; g_{m} \simeq \frac{v_g}{ \pi g_1} \kappa_{m-1}  \;\; (m=1,2,..,N-1)
\]
 for the coupling constants, yielding $g_1 \gg |g_2| \gg |g_3| \gg ...$. These relations readily follow from solving the Hankel system perturbatively under $\kappa_0 \gg |\kappa_1| \gg |\kappa_2| \gg ...$. 
 
 This reverse-engineering procedure thus provides a general and fully constructive route to design a giant atom whose relaxation dynamics exhibits an 
$N-{th}$ order non-Markovian exceptional point at any prescribed location on the real axis.

As an illustrative example, let us provide the full explicit derivation for the case of three contact points $N=3$. Fix $\tau$ , $\varphi=0$ and $s=s_{\mathrm{EP}}<0$, and let calculate the couplings
$\kappa_0$, $ \kappa_1$ and $\kappa_2$ by solving the linear system [Eq.(A5)]
\begin{align}
s_{\mathrm{EP}} + \kappa_0 + \kappa_1 e^{-s_{\mathrm{EP}}\tau} + \kappa_2 e^{-2s_{\mathrm{EP}}\tau} = 0,
\\
1-\,\tau\,\kappa_1  e^{-s_{\mathrm{EP}}\tau}
- 2\tau\,\kappa_2  e^{-2s_{\mathrm{EP}}\tau}
= 0, 
\\
\ \kappa_1 
+ 4 \kappa_2 e^{-s_{\mathrm{EP}}\tau}
=0,
\end{align}
yielding
\begin{eqnarray}
\kappa_0 & = & -s_{\mathrm{EP}}- \frac{3}{2 \tau} \\
\kappa_1 & = & \frac{2}{\tau} e^{ s_{\mathrm{EP}} \tau} \\
\kappa_2 & = & -\frac{1}{2 \tau} e^{2 s_{\mathrm{EP}} \tau}.
\end{eqnarray}
Note that, since $\kappa_0$ should be positive [see Eq.(A2)], from Eq.(A9) it follows that a {\em necessary} condition to obtain an EP is that $\tau$ should be larger than $ 3/  (2|s_{EP}|)$. This shows that an EP strictly requires a sufficiently strong delay, i.e. the EP cannot be observed in the Markovian limit $\tau \rightarrow 0$.
Once the coefficients $\kappa_m$ are known, define
\[
K_m = \frac{v_g}{\pi}\,\kappa_m.
\]
For $N=3$, the relations between physical couplings $g_j$ and $K_m$ are
\begin{align}
K_0 &= g_1^2 + g_2^2 + g_3^2, \label{eq:K0_N3_app}\\
K_1 &= 2(g_1 g_2 + g_2 g_3), \label{eq:K1_N3_app}\\
K_2 &= 2 g_1 g_3. \label{eq:K2_N3_app}
\end{align}
This is a Hankel-type quadratic system, which can be solved as follows.  From \eqref{eq:K2_N3_app},
\[
g_3 = \frac{K_2}{2 g_1}.
\]
Insert into \eqref{eq:K1_N3_app}:
\[
g_2 = \frac{K_1 g_1}{2 g_1^2 + K_2}.
\]
Insert both into \eqref{eq:K0_N3_app} to determine $g_1$:
\begin{equation}
g_1^2 
+
\frac{K_1^2 g_1^2}{(2 g_1^2 + K_2)^2}
+
\frac{K_2^2}{4 g_1^2}
=
K_0.
\end{equation}
This is an algebraic equation, quartic in $X=g_1^2$, which can be written as $F(X)=K_0$, where
\[
F(X) \equiv X+ \frac{K_1^2 X}{(2X+K_2)^2}+\frac{K_2^2}{4X}.
\]
The reverse-engineering procedure yields a physically realizable model provided that the equation $F(X)=K_0$ has a real positive root. 
For large enough delay time $\tau$ such that $ |\tau s_{\mathrm{EP}}| \gg1$, one has $K_0 \gg| K_1| \gg |K_2|$, and a straightforward perturbative analysis indicates that there is always a real positive solution $X=X_0$ to the equation $F(X)=K_0$ with $X_0 \simeq K_0$, i.e. $g_1 \simeq \sqrt{K_0}= \sqrt{v_g \kappa_0 / \pi}$.\\

\end{document}